\documentclass[11pt]{article}
\usepackage{amssymb,amsmath,epsfig}
\usepackage{float}
\allowdisplaybreaks
\def\one{{{{\rm 1} \kern -.19em {\rm l}}}}
\def\C{{{{\rm {\mbox{\small l}}} \kern -.50em {\rm C}}}}
\def\R{{{{\rm l} \kern -.15em {\rm R}}}}
\def\N{{{{\rm l} \kern -.15em {\rm N}}}}
\def\E{{{{\rm l} \kern -.15em {\rm E}}}}
\def\P{{{{\rm l} \kern -.15em {\rm P}}}}
\def\Z{{{{\rm Z} \kern -.35em {\rm Z}}}}
\def\1{{{{\rm 1} \kern -.35em {\rm 1}}}}

\begin{document}
\begin{sloppypar}
\vspace*{0cm}
\begin{center}
{\setlength{\baselineskip}{1.0cm}{ {\Large{\bf
DARBOUX TRANSFORMATIONS FOR DUNKL-SCHR\"ODINGER EQUATIONS WITH ENERGY-DEPENDENT POTENTIAL AND POSITION-DEPENDENT MASS\\}} }}
\vspace*{1.0cm}
{\large{\sc{Axel Schulze-Halberg}}}$^\dagger$ ~~and~~ {\large{\sc{Pinaki Roy}}}$^{\ddagger, \ast}$
\end{center}
\noindent \\
$\dagger~$Department of Mathematics and Actuarial Science and Department of Physics, Indiana University Northwest, 3400 Broadway,
Gary IN 46408, USA, \\E-mail: axgeschu@iun.edu \\ \\
$\ddagger~$Atomic Molecular and Optical Physics Research Group, Advanced Institute of Materials Science, 
Ton Duc Thang University, Ho Chi Minh City, Vietnam \\ \\
$\ast~$Faculty of Applied Sciences, Ton Duc Thang University, Ho Chi Minh City, Vietnam,\\ E-mail: pinaki.roy@tdtu.edu.vn
\vspace*{.5cm}
\begin{abstract} 
\noindent
We construct arbitrary-order Darboux transformations for Schr\"odinger equations with energy-dependent potential and position-dependent mass within the Dunkl formalism. Our construction 
is based on a point transformation that interrelates our equations with the standard Schr\"odinger case. We apply our method to generate several solvable Dunkl-Schr\"odinger 
equations. 

\end{abstract}

 \noindent \\

\section{Introduction}
Dunkl operators \cite{dunkl} are commuting differential-difference operators associated with finite reflection groups. As generalizations of partial derivatives, these 
operators appear in a wide range of mathematical applications, such as Fourier analysis related to root systems \cite{anker}, 
 intertwining operator and angular momentum algebra \cite{feigin1} \cite{feigin2}, integrable Calogero-Moser-Sutherland (CMS) models \cite{diejen} \cite{etingof}, 
harmonic analysis \cite{bouze}, the generation of orthogonal polynomials \cite{luo}, and nonlinear wave equations \cite{mejj}. Another field of application 
arises in quantum physics when conventional derivatives are replaced by Dunkl operators. This replacement can be done in the momentum 
operators, such that the associated Hamiltonian generates governing equations that contain modified derivatives and reflection operators. In the simplest case 
of a one-dimensional nonrelativistic system \cite{chung}, the reflection operator entering in the Schr\"odinger equation coincides with the parity operator. If this equation 
admits solutions that have parity, then for these solutions the parity operator can be replaced by a numerical parameter, which in some cases allows for closed-form 
construction of the latter solutions. Based on this concept, results have been obtained for a variety of systems, for example the three-dimensional Coulomb problem 
\cite{ghazo}, three-dimensional relativistic and nonrelativistic systems with radial symmetry \cite{mota2} \cite{mota3}, and the isotropic two-dimensional Dunkl oscillator 
in the plane \cite{genest1} \cite{genest2}, just to mention a few. While the latter applications encompass several types of governing equations, in the present work we will focus on the 
one-dimensional Schr\"odinger equation that is equipped with a position-dependent mass and an energy-dependent potential. Implementation of these concepts within the Dunkl 
formalism lead to a straightforward modification of the weighted Hilbert space associated with the Hamiltonian \cite{xbat}, being 
very similar to the standard formulation without Dunkl operators \cite{xbatozi}. In order to find cases of our Dunkl-Schr\"odinger equation that admits closed-form solutions, we 
adapt point transformations and Darboux transformations to the present scenario. Both of these techniques have been widely applied to nonrelativistic quantum systems with 
position-dependent mass, for example in the study of finite gap systems \cite{bravo}, the problem of operator ordering in Hamiltonians \cite{karthiga}, and the construction of 
coherent states \cite{dacosta}. In contrast to these and related applications, to the best of our knowledge point transformations and Darboux transformations have not been applied yet 
to Dunkl-Schr\"odinger equations equipped with both a position-dependent mass and an energy-dependent potential. After collecting some basic facts about the Darboux transformation 
in section 2, we outline basic quantum theory in section 3 as it applies to systems with position-dependent mass and energy-dependent potentials. Section 4 is devoted to 
point transformations and their application to a particular system that admits bound states. In section 5 we develop Darboux transformations for the systems under consideration here, 
and we apply both the standard and the confluent Darboux algorithm to generate solvable Dunkl-Schr\"odinger systems that support bound states.

\section{Preliminaries: Darboux transformations}
In order to make this work self-contained, we will now present a brief summary of basic facts about Darboux transformations. Our starting point is the 
pair of Schr\"odinger equations
\begin{eqnarray}
\Phi''(y) + \left[\epsilon - U(y) \right] \Phi(y) &=& 0 \label{sse} \\[1ex]
\hat{\Phi}''(y) + \left[ \epsilon - \hat{U}(y) \right] \hat{\Phi}(y) &=& 0, \label{sset}
\end{eqnarray}
where $\epsilon$ denotes the real-valued stationary energy, the functions $U$, $\hat{U}$ stand for the potentials, and 
$\Phi$, $\hat{\Phi}$ are solutions of their respective equation. In order to establish an interrelation between the two 
equations (\ref{sse}) and (\ref{sset}) by means of a Darboux transformation, we distinguish between the standard algorithm and the 
confluent algorithm.
\begin{itemize}
\item {\bf{Standard algorithm:}} 
We choose solutions (called transformation functions) $u_1,u_2,...,u_n$ of our initial equation (\ref{sse}), that pertain to the 
stationary energies (called transformation energies) $\epsilon_1, \epsilon_2,...,\epsilon_n$, respectively. This means that we have
\begin{eqnarray}
u_j''(y) + \left[ \epsilon_j - U(y) \right] u_j(y) &=& 0,~~~j=1,2,...,n. \nonumber
\end{eqnarray}
Then, the function
\begin{eqnarray}
\hat{\Phi}(y) &=& \frac{W_{u_1,u_2,...,u_{n},\Phi}(y)}{W_{u_1,u_2,...,u_{n}}(y)}, \label{susy1} 
\end{eqnarray}
is a solution of equation (\ref{sset}), provided the transformed potential $\hat{U}$ satisfies the constraint
\begin{eqnarray}
\hat{U}(y) &=& U(y) - 2~\frac{d^2}{dy^2}~\log\left[W_{u_1,u_2,...,u_{n}}(y)\right]. \label{susy2}
\end{eqnarray}
We refer to (\ref{susy1}) as Darboux transformation of order $n$. 
\item {\bf{Confluent algorithm:}}
In contrast to the previous case, we have a single transformation energy $\epsilon_1$ only, and the transformation functions 
obey the following system
\begin{eqnarray}
u_1''(y)+\left[\epsilon_1-U(y) \right] u_1(y) &=& 0 \label{eq1} \\[1ex]
u_j''(y)+\left[\epsilon_1-U(y)\right]~u_j(y) &=&- u_{j-1}(y),~~~j=2,...,n. \label{eqn}
\end{eqnarray}
The expressions for the transformed solution (\ref{susy1}) and potential (\ref{susy2}) remain the same as before.
\end{itemize}

\section{The Dunkl-Schr\"odinger quantum system}
Our goal for this section is to set up a Hamiltonian for position-dependent mass within the Dunkl formalism. In addition, we will 
define a domain for our Hamiltonian, such that it renders hermitian within a suitable Hilbert space. To this end, let us first 
introduce the simplest case of a Dunkl operator \cite{dunkl} in the form
\begin{eqnarray}
D_x &=& \frac{d}{dx}+\frac{\nu}{x}-\frac{\nu}{x}~{\cal R}, \label{dop}
\end{eqnarray}
where $\nu$ is a real number, and ${\cal R}$ stands for the reflection or parity operator. While 
in the scenario of constant mass we have $\nu > -1/2$, in the presence of a position-dependent mass admissible values for the parameter 
$\nu$ are determined through the Hilbert space that we will define below. Next, we use our operator (\ref{dop}) to 
build the Hamiltonian in the form
\begin{eqnarray}
H &=& -D_x~\frac{1}{2~m(x)}~D_x +V_E(x), \label{dham}
\end{eqnarray}
introducing a position-dependent mass $m$, and an energy-dependent potential $V_E$, both of which are 
assumed to be sufficiently smooth functions that are defined on the whole real line except possibly at the origin. 
Let us simplify subsequent calculations by requiring the mass to have parity. More precisely, we set
\begin{eqnarray}
{\cal R} ~m(x) &=& \mu~m(x), \label{rm}
\end{eqnarray}
such that an odd mass is represented by $\mu=-1$, while an even mass corresponds to $\mu=1$. In the same way we describe the 
action of the reflection operator on a function $\Psi$ that the Hamiltonian (\ref{dham}) acts on. We set
\begin{eqnarray}
{\cal R} ~\Psi(x) &=& \delta~\Psi(x), \label{delta}
\end{eqnarray}
where in the odd case we have 
$\delta=-1$ and in the even case $\delta=1$ holds. Now, the Hamiltonian (\ref{dham}) is hermitian in the weighted Hilbert 
space $L_w^2(\mathbb{R})$, where the weight function $w$ is given by
\begin{eqnarray}
w(x) &=& |x|^{2 \nu-\delta \nu+\frac{\delta \nu}{\mu}}. \label{w}
\end{eqnarray}
The Dunkl-Schr\"odinger equation is obtained from our Hamiltonian (\ref{dham}) in the form
\begin{eqnarray}
D_x~\frac{1}{2~m(x)}~D_x~\Psi(x) + \left[E-V_E(x) \right] \Psi(x) &=& 0. \label{sse0}
\end{eqnarray}
where $\Psi$ represents a solution to the equation. Upon taking into account our weight function (\ref{w}) and the energy 
dependence of our potential, the modified probability density $P$ for a solution $\Psi$ of (\ref{sse0}) in 
$L_w^2(\mathbb{R})$ is given by
\begin{eqnarray}
P_\Psi(x) &=& \Psi(x)^\ast \Psi(x)~ |x|^{2 \nu-\delta \nu+\frac{\delta \nu}{\mu}}
\left[1-\frac{\partial V_E(x)}{\partial E}\right]. \label{prob}
\end{eqnarray}
This defines the associated modified norm as
\begin{eqnarray}
N(\Psi) &=& \int\limits_{\mathbb{R}} P_\Psi(x)~dx ~=~ \int\limits_{\mathbb{R}} \Psi(x)^\ast \Psi(x)~ |x|^{2 \nu-\delta \nu+\frac{\delta \nu}{\mu}}
\left[1-\frac{\partial V_E(x)}{\partial E}\right] dx. \label{norm}
\end{eqnarray}
If the potential $V_E$ does not depend on the energy, the term in square brackets equals one. In this case we must require 
\begin{eqnarray}
2 \nu-\delta \nu+\frac{\delta \nu}{\mu} &>& -1, \label{nures}
\end{eqnarray}
such that the term $\sim |x|$ does not contribute singularities to the integrand. For constant mass we have $\mu=1$, implying the known 
restriction $\nu>-1/2$. If the potential $V_E$ is energy-dependent, then the restriction (\ref{nures}) can be dropped, since 
the term in square brackets can contribute singularities or remove them, depending on the behavior of the potential $V_E$. In such a 
situation we must find restrictions for the parameter $\nu$ on a case-by-case basis. This is illustrated in our application section 4.2. 
Let us now write (\ref{sse0}) in expanded form by substituting (\ref{dop}) and by incorporation of (\ref{delta}). This gives
\begin{eqnarray}
& & \hspace{-1cm}\frac{1}{2~m(x)}~\Psi''(x)+
\Bigg[\hspace{-.1cm}
-\frac{m'(x)}{2~m(x)^2}+\frac{\nu}{m(x)~x}-\frac{\nu~\delta}{2~m(x)~x}+
\frac{\nu~\delta}{2~\mu~m(x)~x}
\Bigg]~\Psi'(x)+ \nonumber \\[1ex]
&+&
\Bigg[-\frac{\nu}{2~m(x)~x^2}-
\frac{\nu~m'(x)}{2~m(x)^2~x}+\frac{\nu~\delta}{2~m(x)~x^2}+\frac{\nu~\delta~m'(x)}{2~m(x)^2~x}+\frac{\nu^2}{2~m(x)~x^2}-
\nonumber \\[1ex]
&-&\frac{\nu^2~\delta}{2~m(x)~x^2}+\frac{\nu^2~\delta}{2~\mu~m(x)~x^2}-\frac{\nu^2}{2~\mu~m(x)~x^2} +E-V_E(x)\Bigg]~\Psi(x) ~=~ 0. \label{dssegen}
\end{eqnarray}
If we distinguish the parameter values for $\delta$ and $\mu$, we obtain four special cases of this equation. Let us show the 
two cases that arise from odd and even position-dependent mass functions. We will study the two remaining cases in detail below. 
For an odd-parity mass ($\mu=-1$) our equation takes the form
\begin{eqnarray}
& & \hspace{-1cm} \frac{1}{2~m(x)}~\Psi''(x)+
\Bigg[\hspace{-.1cm}
-\frac{m'(x)}{2~m(x)^2}+\frac{\nu}{m(x)~x}-\frac{\nu~\delta}{m(x)~x}
\Bigg] \Psi'(x)+ \Bigg[
-\frac{\nu}{2~m(x)~x^2}-
\frac{\nu~m'(x)}{2~m(x)^2~x}+ \nonumber \\[1ex]
& &\hspace{-1cm} +~\frac{\nu~\delta}{2~m(x)~x^2}+\frac{\nu~\delta~m'(x)}{2~m(x)^2~x}+
\frac{\nu^2}{m(x)~x^2}-\frac{\nu^2~\delta}{m(x)~x^2}+E-V_E(x) \Bigg] \Psi(x) ~=~ 0. \label{dsse0}
\end{eqnarray}
If our mass function has even parity ($\mu=1)$, then the Dunkl-Schr\"odinger equation (\ref{dssegen}) reads
\begin{eqnarray}
& & \hspace{-1cm} \frac{1}{2~m(x)}~\Psi''(x)+
\Bigg[\hspace{-.1cm}
-\frac{m'(x)}{2~m(x)^2}+\frac{\nu}{m(x)~x}
\Bigg] \Psi'(x)+\Bigg[-
\frac{\nu}{2~m(x)~x^2}-
\frac{\nu~m'(x)}{2~m(x)^2~x}+\frac{\nu~\delta}{2~m(x)~x^2}+ \nonumber \\[1ex]
& &\hspace{-1cm} +~\frac{\nu~\delta~m'(x)}{2~m(x)^2~x}+E-V_E(x) \Bigg] \Psi(x)~=~0. \label{dsse1}
\end{eqnarray}
A particular case of this equation is the scenario of constant mass. We obtain for $m=1/2$:
\begin{eqnarray}
\Psi''(x)+\frac{2~\nu}{x}
~\Psi'(x)+\Bigg[\frac{\nu~\delta-\nu}{x^2}+E-V_E(x) \Bigg] \Psi(x)~=~0. \label{mconst}
\end{eqnarray}
Note that the mass being constant implies $\mu=1$.

\section{Point transformations}
The first method for solving the Dunkl-Schr\"odinger equation (\ref{dssegen}) consists in a point transformation that takes it 
into conventional Schr\"odinger form by gauging away the first-derivative term $\sim \Psi'$. Appropriate choices for the 
free parameter functions then allow for the construction of a solvable case. 
\subsection{Transformation to standard form}
We apply the following point transformation
\begin{eqnarray}
\Phi(y) &=& \sqrt{\frac{1}{m[x(y)]~x'(y)}}~x(y)^{\nu-\frac{\delta \nu}{2}+\frac{\delta \nu}{2 \mu}}~\Psi[x(y)]. \label{pct}
\end{eqnarray}
The coordinate change $x=x(y)$ remains arbitrary for now, except that it must have an inverse. It is sufficient if the inverse exists 
on either the negative or the positive axis, since we are considering solutions with parity only. The specfic form of 
(\ref{pct}) ensures that the coefficient of the term $\sim \Psi'$ vanishes. Substitution of (\ref{pct}) renders (\ref{dssegen}) in the form
\begin{eqnarray}
\Phi''(y) \hspace{-.2cm} &+& \hspace{-.2cm}
\Bigg\{
2~E~m[x(y)]~[x'(y)]^2-2~m[x(y)]~V_E[x(y)]~[x'(y)]^2+\frac{\delta~\nu~[x'(y)]^2}{2~x(y)^2}+\frac{\delta~\nu~[x'(y)]^2}{2~\mu~x(y)^2}-
\nonumber \\[1ex]
&-&
\frac{\nu^2~[x'(y)]^2}{2~x(y)^2}-\frac{\nu^2~[x'(y)]^2}{2~\mu~x(y)^2}+\frac{\delta~\nu~m'[x(y)]~[x'(y)]^2}{2~m[x(y)]~x(y)}+
\frac{\delta~\nu~m'[x(y)]~[x'(y)]^2}{2~\mu~m[x(y)]~x(y)}-
\nonumber \\[1ex]
&-&
\frac{3~\{m'[x(y)]\}^2~[x'(y)]^2}{4~m[x(y)]}+\frac{[x'(y)]^2~m''[x(y)]}{2~m[x(y)]}
-\frac{3~[x''(y)]^2}{4~[x'(y)]^2}+\frac{x'''(y)}{2~x'(y)}
\Bigg\}~\Phi(y) ~=~0. \label{dsse}
\end{eqnarray}
Let us now recall that the conventional Schr\"odinger equation with energy-dependent potential $U_E$ reads
\begin{eqnarray}
\Phi''(y)+\left[E-U_E(y)\right] \Phi(y) &=& 0. \label{sse00}
\end{eqnarray}
Hence, for the transformed Dunkl-Schr\"odinger equation (\ref{dsse}) to take the form (\ref{sse00}), the following condition must be 
imposed on the potential $U_E$:
\begin{eqnarray}
U_E(y) &=& E-
2~E~m[x(y)]~[x'(y)]^2+2~m[x(y)]~V_E[x(y)]~[x'(y)]^2-\frac{\delta~\nu~[x'(y)]^2}{2~x(y)^2}- \nonumber \\[1ex] 
& & \hspace{-1.5cm}-~\frac{\delta~\nu~[x'(y)]^2}{2~\mu~x(y)^2}+
\frac{\nu^2~[x'(y)]^2}{2~x(y)^2}+\frac{\nu^2~[x'(y)]^2}{2~\mu~x(y)^2}-\frac{\delta~\nu~m'[x(y)]~[x'(y)]^2}{2~m[x(y)]~x(y)}-
\frac{\delta~\nu~m'[x(y)]~[x'(y)]^2}{2~\mu~m[x(y)]~x(y)}+
\nonumber \\[1ex]
& & \hspace{-1.5cm}+~
\frac{3~\{m'[x(y)]\}^2~[x'(y)]^2}{4~m[x(y)]}-\frac{[x'(y)]^2~m''[x(y)]}{2~m[x(y)]}
+\frac{3~[x''(y)]^2}{4~[x'(y)]^2}+\frac{x'''(y)}{2~x'(y)}. \label{con}
\end{eqnarray}
Note that this condition arises simply by comparing the coefficients of $\Phi$ in the two equations (\ref{dsse}) and (\ref{sse00}). In the 
special case of constant mass $m=1/2$ and $\mu=1$, the condition (\ref{con}) reads
\begin{eqnarray}
U_E(y) \hspace{-.2cm} &=& \hspace{-.2cm} E-E~[x'(y)]^2+V_E[x(y)]~[x'(y)]^2-\frac{\delta~\nu~[x'(y)]^2}{x(y)^2}+\frac{\nu^2~[x'(y)]^2}{x(y)^2}+\frac{3~[x''(y)]^2}{4~[x'(y)]^2}
-\frac{x'''(y)}{2~x'(y)}. \nonumber \\ \label{con0}
\end{eqnarray}
Since our conditions (\ref{con}) and (\ref{con0}) contain three and two free parameter functions, respectively, the same function 
$U_E$ can be obtained by different choices for the latter parameter functions. As such, each $U_E$ gives rise to a class of 
Dunkl-Schr\"odinger systems with energy-dependent potential and position-dependent mass. We will comment further on this 
property of (\ref{con}) and (\ref{con0}) in our application section 5.3. 
Let us now present examples for the construction of a Dunkl-Schr\"odinger system with energy-dependent potential by means of 
point transformations.

\subsection{Application}
\begin{figure}[H]
\begin{center}
\epsfig{file=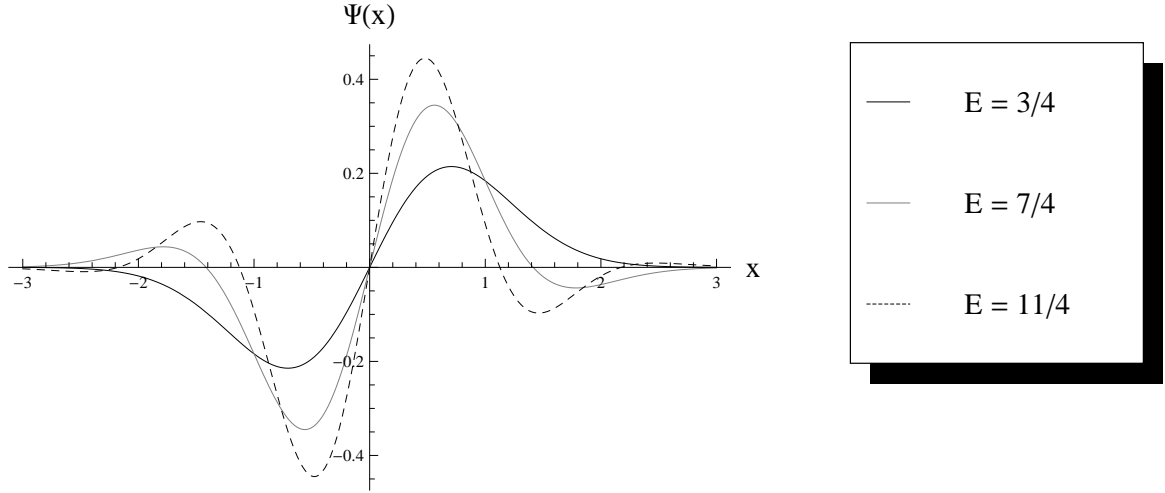,width=16cm}
\caption{Graphs of the odd solutions (\ref{psix0})-(\ref{psix2}).} 
\label{fig0}
\end{center}
\end{figure}

\begin{figure}[H]
\begin{center}
\epsfig{file=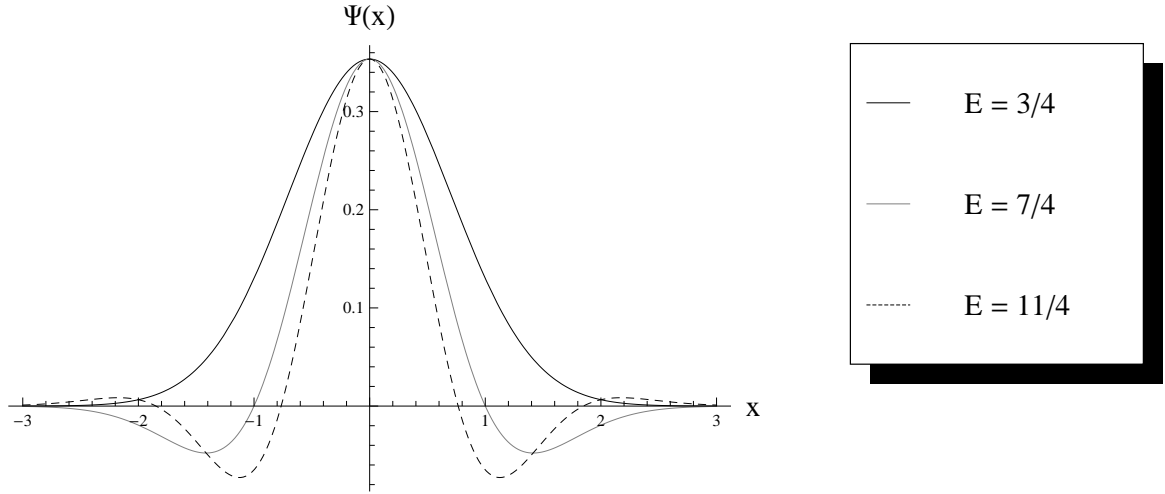,width=16cm}
\caption{Graphs of the even solutions (\ref{psix3})-(\ref{psix5}).} 
\label{fig1}
\end{center}
\end{figure}

\begin{figure}[H]
\begin{center}
\epsfig{file=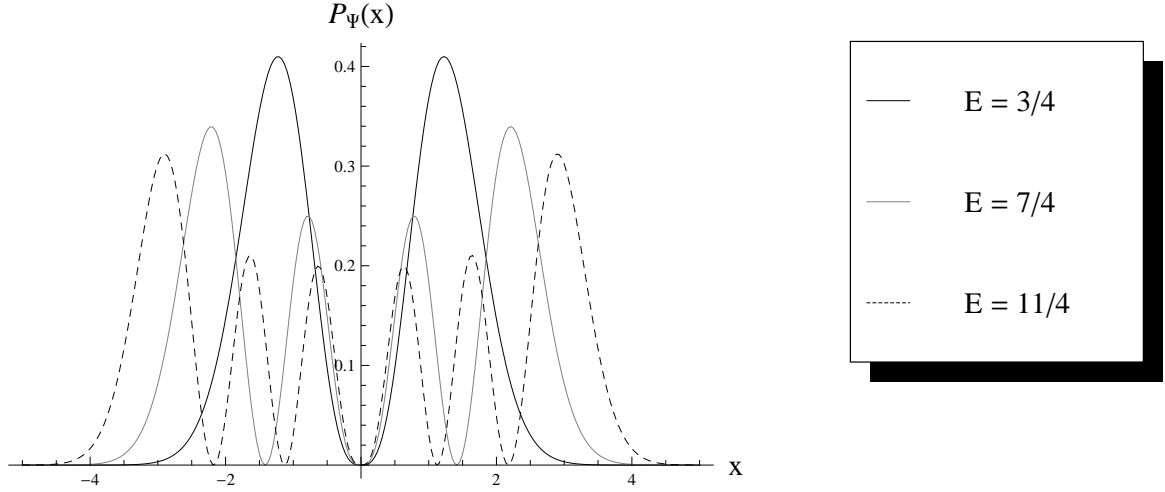,width=16cm}
\caption{Graphs of normalized probability densities pertaining to the odd solutions (\ref{psix0})-(\ref{psix2}).} 
\label{fig2}
\end{center}
\end{figure}

Our starting point is the conventional Schr\"odinger equation (\ref{sse}). If we can generate a potential $U_E$ that satisfies the condition 
(\ref{con}), such that (\ref{sse}) is solvable, then our point transformation (\ref{pct}) allows for the construction of an associated 
Dunkl-Schr\"odinger equation that is also solvable. Let us make the following choices for the position-dependent mass and the 
potential in the Dunkl-Schr\"odinger system that we want to construct:
\begin{eqnarray}
m(x) &=& p~\exp\left(-q~x^2 \right) \label{massgen} \\[1ex] 
V_E(x) &=& E-2~p~E~\exp\left(q~x^2 \right)-\frac{p}{2}~\exp\left(q~x^2 \right) \label{vgen} \\[1ex]
x(y) &=& \sqrt{y}, \label{set0}
\end{eqnarray}
where $p$ and $q$ stand for arbitrary real numbers. We will use particular cases of our specific mass profile (\ref{massgen}) in the 
subsequent sections. Note that our change of coordinate is defined and invertible on the positive half-axis, which is sufficient for our purposes. 
Furthermore, we chose the particular change of coordinate (\ref{set0}), such that the 
conventional Schr\"odinger 
equation (\ref{sse}) solvable. Substitution of (\ref{set0}) into the potential condition (\ref{con}) yields
\begin{eqnarray}
U_E(y) &=& E+\frac{q^2}{4}-\left(-\frac{q}{4}+\frac{p^2}{4}+p^2~E -\frac{q~\delta~\nu}{2}
\right)\frac{1}{y}-\left(\frac{3}{16}+\frac{\delta~\nu}{4}-\frac{\nu^2}{4} 
\right)\frac{1}{y^2}. \nonumber \\ \label{uegen}
\end{eqnarray}
In order to keep our calculations simple and transparent, let us take a special case of the mass function (\ref{massgen}) by 
applying the parameter setting $p=q=1$. Then (\ref{uegen}) becomes
\begin{eqnarray}
U_E(y) &=& E+\frac{1}{4}-\left(E-\frac{\delta~\nu}{2}\right)\frac{1}{y}-\left(\frac{3}{16}+\frac{\delta~\nu}{4}-\frac{\nu^2}{4} 
\right)\frac{1}{y^2}. \nonumber
\end{eqnarray}
The corresponding equation (\ref{sse}) then takes the form
\begin{eqnarray}
\Phi''(y)+\left[ -\frac{1}{4}+\left(E-\frac{\delta~\nu}{2}\right)\frac{1}{y}+\left(\frac{3}{16}+\frac{\delta~\nu}{4}-\frac{\nu^2}{4} 
\right)\frac{1}{y^2}
\right] \Phi(y) &=& 0. 
\end{eqnarray}
The general solution to this equation can be given in terms of hypergeometric functions. We have
\begin{eqnarray}
\Phi_{\tiny{\mbox{gen}}}(y) &=& \exp\left(-\frac{y}{2} \right) y^{\frac{1}{2}+\frac{1}{4} \sqrt{1-4 \delta \nu+4 \nu^2}} \times \nonumber \\[1ex]
&\times&
\Bigg[c_1~{}_1F_1\left(\frac{1}{2}-E+\frac{\delta~\nu}{2}+\frac{1}{4}~ \sqrt{1-4 ~\delta ~\nu+4 ~\nu^2},
1+\frac{1}{2}~ \sqrt{1-4 ~\delta~ \nu+4 ~\nu^2},y\right)+ \nonumber \\[1ex]
&+& c_2~U\left(\frac{1}{2}-E+\frac{\delta~\nu}{2}+\frac{1}{4}~ \sqrt{1-4 ~\delta ~\nu+4 ~\nu^2},
1+\frac{1}{2}~ \sqrt{1-4 ~\delta~ \nu+4 ~\nu^2},y\right) \Bigg], \label{phigen}
\end{eqnarray}
where $c_1$ and $c_2$ represent arbitrary constants. Furthermore, ${}_1F_1$ and $U$ stand for the confluent 
hypergeometric functions of first and second kind, respectively \cite{abram}. 
Since we are aiming for the construction of bound states, we must work with a solution that is bounded on the whole 
positive half-axis. This condition is not satisfied by the function $U$, such that we can set $c_1=1$ and $c_2=0$ in 
(\ref{phigen}). As a result, we obtain the particular solution 
\begin{eqnarray}
\Phi(y) &=& \exp\left(-\frac{y}{2} \right) y^{\frac{1}{2}+\frac{1}{4} \sqrt{1-4 \delta \nu+4 \nu^2}} \times \nonumber \\[1ex]
&\times&
{}_1F_1\left(\frac{1}{2}-E+\frac{\delta~\nu}{2}+\frac{1}{4}~ \sqrt{1-4 ~\delta ~\nu+4 ~\nu^2},
1+\frac{1}{2}~ \sqrt{1-4 ~\delta~ \nu+4 ~\nu^2},y\right). \nonumber
\end{eqnarray}
Upon reversing our point transformation (\ref{pct}), 
we obtain the function $\Psi$ in the form
\begin{eqnarray}
\Psi(x) &=& x^{\frac{1}{2}-\nu+\frac{1}{2} \sqrt{1-4 \delta \nu+4 \nu^2}} \times \nonumber \\[1ex]
&\times&
{}_1F_1\left(\frac{1}{2}+E-\frac{\delta~\nu}{2}+\frac{1}{4}~ \sqrt{1-4 ~\delta ~\nu+4 ~\nu^2},
1+\frac{1}{2}~ \sqrt{1-4 ~\delta~ \nu+4 ~\nu^2},-x^2\right). \label{psi00}
\end{eqnarray}
It is convenient to rewrite the confluent hypergeometric function by means of an identity \cite{wolfram} as follows
\begin{eqnarray}
\Psi(x) &=& \exp\left(-x^2 \right) x^{\frac{1}{2}-\nu+\frac{1}{2} \sqrt{1-4 \delta \nu+4 \nu^2}} \times \nonumber \\[1ex]
&\times&
{}_1F_1\left(\frac{1}{2}-E+\frac{\delta~\nu}{2}+\frac{1}{4}~ \sqrt{1-4 ~\delta ~\nu+4 ~\nu^2},
1+\frac{1}{2}~ \sqrt{1-4 ~\delta~ \nu+4 ~\nu^2},x^2\right). \label{psi0}
\end{eqnarray}
Since the point transformation (\ref{pct}) interrelates the Dunkl-Schr\"odinger equation (\ref{dssegen}) and its 
conventional counterpart (\ref{sse}), the function (\ref{psi0}) must solve (\ref{dssegen}). 
Upon plugging the settings (\ref{massgen}) and (\ref{vgen}) for $p=q=1$ into (\ref{dssegen}), we obtain the latter equation as
\begin{eqnarray}
& &\frac{1}{2}~\Psi''(x)+\left(x+\frac{\nu}{x} \right)\Psi'(x)+
\left(
\frac{1}{2}+2~E+\nu-\frac{\nu}{2~x^2}
-\delta~\nu+\frac{\delta~\nu}{2~x^2}
\right) \Psi(x) ~=~ 0. \nonumber
\end{eqnarray}
Note that even if (\ref{psi0}) is a solution to the latter equation, it does not automatically solve the initial Dunkl-Schr\"odinger equation 
(\ref{sse0}), unless it has the correct parity behavior given by (\ref{delta}). In order to investigate this behavior, we note that both the 
exponential and the confluent hypergeometric function are even, such that the overall behavior depends on the exponent of the 
monomial term. Let us distinguish the two cases of $\delta=-1$ and $\delta=1$. In the first of these cases the exponent reads
\begin{eqnarray}
\frac{1}{2}-\nu+\frac{1}{2}~\sqrt{1+4~ \nu+4~ \nu^2} &=& \frac{1}{2}-\nu+\frac{1}{2}~|1+2~\nu|. \label{parm}
\end{eqnarray} 
At this point we cannot proceed until we know the admissible values for the parameter $\nu$. According to section 3, due to the 
energy dependence of our potential $V_E$, we must find these admissible values from the behavior of the term 
$\partial V_E / \partial E$ that enters in (\ref{norm}). Substitution of (\ref{set0}) yields
\begin{eqnarray}
1-\frac{\partial V_E(x)}{\partial E} &=& 2~\exp\left(x^2 \right), \label{dv}
\end{eqnarray}
which gives the standard restriction $\nu>-1/2$ because the confluent hypergeometric function is regular at the origin in the 
cases of bound states that we are interested in. Upon using our restriction in (\ref{parm}), we obtain
\begin{eqnarray}
\frac{1}{2}-\nu+\frac{1}{2}~|1+2~\nu| ~=~ 
\frac{1}{2}-\nu+\frac{1}{2}~(1+2~\nu) ~=~ 1. \nonumber
\end{eqnarray}
Since the exponent is an odd number, the function $(\ref{psi0})$ has odd parity, and 
as such solves the initial Dunkl-Schr\"odinger equation. Next we consider the case $\delta=1$, the exponent of the middle term on 
the right side of (\ref{psi0}) takes the form
\begin{eqnarray}
\frac{1}{2}-\nu+\frac{1}{2}~\sqrt{1-4~ \nu+4~ \nu^2} &=& \frac{1}{2}-\nu+\frac{1}{2}~|1-2~\nu| \nonumber \\[1ex]
&=& \left\{
\begin{array}{lll}
1-2~\nu & \mbox{if} & \nu < \frac{1}{2} \\[1ex]
0 & \mbox{if} & \nu \geq \frac{1}{2}
\end{array}
\right\}. \nonumber
\end{eqnarray}
Hence, if $\delta=1$ and $\nu \geq 1/2$, the function (\ref{psi0}) has even parity, such that it solves our initial 
Dunkl-Schr\"odinger equation. If $\nu < 1/2$, then the exponent depends on $\nu$, resulting in the condition
\begin{eqnarray}
1-2~\nu &=& 2~k, \label{nucon}
\end{eqnarray}
for an arbitrary integer $k$. If we take into account that $\nu > -1/2$ due to (\ref{dv}), there is no solution of (\ref{nucon}). 
Hence, in the present case we have the overall constraint
\begin{eqnarray}
\nu &\geq& \frac{1}{2}. \nonumber
\end{eqnarray}
Now, 
since we want to construct solutions of bound state type, we restrict the stationary energy, such 
that the confluent hypergeometric function in (\ref{psi0}) degenerates to a polynomial. This restriction is given by 
equating the first argument of the latter function to a nonpositive integer, leading to
\begin{eqnarray}
E &=& n+\frac{1+\delta~\nu}{2}+\frac{1}{4}~\sqrt{1-4~\delta~\nu+4~\nu^2}, \label{ene0}
\end{eqnarray}
where $n$ is a nonnegative integer. Let us state examples of our solutions (\ref{psi0}) for the parameter setting $\nu=1/2$, 
$\delta =-1$ (odd parity), and the parameter $n$ as given in (\ref{ene0}). We find
\begin{eqnarray}
\Psi(x)_{n=0} &=& \exp\left(-x^2 \right) x  \label{psix0} \\[1ex]
\Psi(x)_{n=1} &=&  \exp\left(-x^2 \right) x \left(1-\frac{x^2}{2} \right) \label{psix1} \\[1ex]
\Psi(x)_{n=2} &=&  \exp\left(-x^2 \right) x \left(1-x^2+\frac{x^4}{2} \right). \label{psix2}
\end{eqnarray}
The graphs of these solutions are shown in figure \ref{fig0}. Now we state the corresponding solutions for the case of even parity 
($\delta=1$), leaving the remaining parameters unchanged. This gives
\begin{eqnarray}
\Psi(x)_{n=0} &=&  \exp\left(-x^2 \right)  \label{psix3} \\[1ex]
\Psi(x)_{n=1} &=&  \exp\left(-x^2 \right) \left(1-x^2 \right) \label{psix4} \\[1ex]
\Psi(x)_{n=2} &=&  \exp\left(-x^2 \right) \left(1-2~x^2+\frac{x^4}{2} \right). \label{psix5}
\end{eqnarray}
Graphs of these solutions are displayed in figure \ref{fig1}. Next, we verify normalizability of our solution (\ref{psi0}) with respect to the modified norm (\ref{norm}). Taking into account that 
(\ref{dv}) is a positive function, and implementing the present settings (\ref{set0}) and $\mu=1$, the probability density and the 
associated norm read
\begin{eqnarray}
P_\Psi(x) &=& \Psi(x)^\ast \Psi(x)~ |x|^{2 \nu}\exp\left(x^2 \right) \nonumber \\[1ex]
N(\Psi) &=& 2~\int\limits_{\mathbb{R}} P_\Psi(x)~ dx, \label{norm0}
\end{eqnarray}
observe that $\delta$ does not enter in the norm due to the mass having even parity. The integral (\ref{norm0}) exists for 
(\ref{psi0}) under the energy restriction (\ref{ene0}), such that the latter solution is normalizable, and thus represents bound states. 
Figure \ref{fig2} shows normalized probability densities associated with the solutions (\ref{psix0})-(\ref{psix2}) and 
(\ref{psix3})-(\ref{psix5}), respectively. 

\section{Darboux transformations}
Besides point transformations that we studied in the previous section, a further method for generating solvable equations 
of Schr\"odinger form is given by Darboux transformations. These cannot be applied to the Dunkl-Schr\"odinger equation (\ref{sse0}) 
directly due to its form, such that we first need to convert it suitably. More precisely, we will use the point transformation (\ref{pct}) 
for this task. In addition, we must distinguish between the cases of odd-parity mass and even-parity mass.

\subsection{Odd-parity mass functions} 
Before we start the construction of our Darboux transformation, let us point out that odd-parity mass functions must be negative 
either on the positive or on the negative axis. Even though there are some applications for negative mass functions, 
to the best of our knowledge they are generally considered non-physical. For this reason we will develop the Darboux transformation, 
but omit to state an example for it. Now, if the mass has odd parity, we have $\mu=-1$ according to (\ref{rm}), such that the Dunkl-Schr\"odinger equation takes the form 
(\ref{dsse0}) after introduction of the parameter $\delta$ in (\ref{delta}). Next, we apply our point transformation (\ref{pct}), 
such that the resulting equation is given by (\ref{dsse}) for $\mu=-1$. In explicit form we have
\begin{eqnarray}
\Phi''(y) \hspace{-.2cm} &+& \hspace{-.2cm}
\Bigg\{
2~E~m[x(y)]~[x'(y)]^2-2~m[x(y)]~V_E[x(y)]~[x'(y)]^2-
\frac{3~\{m'[x(y)]\}^2~[x'(y)]^2}{4~m[x(y)]}+ \nonumber \\[1ex]
&+& \frac{[x'(y)]^2~m''[x(y)]}{2~m[x(y)]}
-\frac{3~[x''(y)]^2}{4~[x'(y)]^2}+\frac{x'''(y)}{2~x'(y)}
\Bigg\}~\Phi(y) ~=~0. \label{dsseodd}
\end{eqnarray}
Inspection of this equation shows that the parameters $\nu$ and $\delta$ from the Dunkl operator (\ref{dop}) are not present 
anymore. In other words, the Dunkl scenario enters in our system entirely by means of the point transformation (\ref{pct}), 
provided the mass function has odd parity. Now, in order to apply a Darboux transformation to (\ref{dsseodd}), we must 
determine a transformation parameter that plays the role of the stationary energy in the conventional Schr\"odinger case. 
We choose this parameter to be $E$, which implies that its coefficient must be equal to one. The condition reads
\begin{eqnarray}
2~m[x(y)]~[x'(y)]^2 &=& 1. \label{mx}
\end{eqnarray}
Implementation of this constraint in (\ref{dsseodd}) renders our equation in the form
\begin{eqnarray}
\Phi''(y)+\left\{E-V_E[x(y)]-\frac{\{m'[x(y)]\}^2}{8~m[x(y)]^3}
\right\} \Phi(y) &=& 0. \label{dsseodd1}
\end{eqnarray}
We can now apply a Darboux transformation to (\ref{dsseodd1}). To this end, we use our formulas (\ref{susy1}) and (\ref{susy2}), 
where the potential $U$ must be replaced by $V_E$. We then have for the transformed solution and the associated potential
\begin{eqnarray}
\hat{\Phi}(y) ~=~ \frac{W_{u_1,...,u_n,\Phi}(y)}{W_{u_1,...,u_n}(y)} \qquad \qquad \qquad 
\hat{V}_E[x(y)] ~=~ V_E[x(y)]- 2~\frac{d^2}{dy^2}~\log\left[
W_{u_1,...,u_n}(y)
\right], \nonumber
\end{eqnarray}
which enter in the transformed Schr\"odinger equation (\ref{sset}) that is the partner of (\ref{dsseodd1}), and 
reads in the present case
\begin{eqnarray}
\hat{\Phi}''(y)+\left\{E-\hat{V}_E[x(y)]-\frac{\{m'[x(y)]\}^2}{8~m[x(y)]^3}
\right\} \hat{\Phi}(y) &=& 0. \nonumber
\end{eqnarray}
The next step consists in reverting the point transformation (\ref{pct}) for reinstating the Dunkl-Schr\"odinger form. This 
reversion is achieved by setting
\begin{eqnarray}
\hat{\Psi}(x) &=& \sqrt{\frac{m(x)}{y'(x)}}~x^{-\nu+\delta \nu}~\hat{\Phi}[y(x)], \nonumber
\end{eqnarray}
where $y=y(x)$ is the inverse function of the initial coordinate change $x=x(y)$. Now, 
the function $\hat{\Psi}$ solves the transformed counterpart of (\ref{dsse0}) that is given by
\begin{eqnarray}
& & \hspace{-1cm} \frac{1}{2~m(x)}~\hat{\Psi}''(x)+
\Bigg[\hspace{-.1cm}
-\frac{m'(x)}{2~m(x)^2}+\frac{\nu}{m(x)~x}-\frac{\nu~\delta}{m(x)~x}
\Bigg] \hat{\Psi}'(x)+ \Bigg[
-\frac{\nu}{2~m(x)~x^2}-
\frac{\nu~m'(x)}{2~m(x)^2~x}+ \nonumber \\[1ex]
& &\hspace{-1cm} +~\frac{\nu~\delta}{2~m(x)~x^2}+\frac{\nu~\delta~m'(x)}{2~m(x)^2~x}+
\frac{\nu^2}{m(x)~x^2}-\frac{\nu^2~\delta}{m(x)~x^2}+E-\hat{V}_E(x) \Bigg] \hat{\Psi}(x) ~=~ 0. \label{dsse0t}
\end{eqnarray}
Let us point out that any solution $\hat{\Psi}$ of this equation only solves the transformed Dunkl-Schr\"odinger equation
\begin{eqnarray}
D_x~\frac{1}{2~m(x)}~D_x~\hat{\Psi}(x) + \left[E-\hat{V}_E(x) \right] \hat{\Psi}(x) &=& 0, \label{sse0t}
\end{eqnarray}
if $\hat{\Psi}$ has the correct parity given by the parameter $\delta$ in (\ref{dsse0t}). In summary, we have 
established Darboux transformations between the Dunkl-Schr\"odinger partner equations (\ref{dsse0}) and 
(\ref{dsse0t}).

\subsection{Even-parity mass functions} 
As in the previous case, our starting point is the Dunkl-Schr\"odinger equation (\ref{sse0}) that we 
consider in the form (\ref{dsse1}) for the parameter setting $\mu=1$. Application of our point transformation (\ref{pct}) yields 
\begin{eqnarray}
\Phi''(y) \hspace{-.2cm} &+& \hspace{-.2cm}
\Bigg\{
2~E~m[x(y)]~[x'(y)]^2-2~m[x(y)]~V_E[x(y)]~[x'(y)]^2+\nonumber \\[1ex]
&+&
\frac{\delta~\nu~[x'(y)]^2}{x(y)^2}-
\frac{\nu^2~[x'(y)]^2}{x(y)^2}+\frac{\delta~\nu~m'[x(y)]~[x'(y)]^2}{m[x(y)]~x(y)}-
\nonumber \\[1ex]
&-&
\frac{3~\{m'[x(y)]\}^2~[x'(y)]^2}{4~m[x(y)]}+\frac{[x'(y)]^2~m''[x(y)]}{2~m[x(y)]}
-\frac{3~[x''(y)]^2}{4~[x'(y)]^2}+\frac{x'''(y)}{2~x'(y)}
\Bigg\}~\Phi(y) ~=~0. \nonumber \\ \label{dsse1even}
\end{eqnarray}
In contrast to the previously considered case of odd-parity mass, this time our equation contains both parameters $\nu$ and 
$\delta$ from the Dunkl operator (\ref{dop}). While so far we have followed the same approach as for the odd-parity mass function, 
at this point we need to proceed differently. The reason lies in the presence of the Dunkl parameters $\nu$ and $\delta$ in 
our equation (\ref{dsse1even}). More precisely, if we used the setting (\ref{mx}) and took the stationary energy $E$ as our 
parameter for the Darboux transformation, then the transformed potential would depend on $\nu$ and $\delta$, which is not 
desirable. Therefore, we would have to assign numerical values to these Dunkl parameters, which would constrain parity of 
the solution to (\ref{dsse1even}). Furthermore, we would need to introduce new constants that play the role of the Dunkl 
parameters, such that the Darboux transformed counterpart of (\ref{dsse1even}) can be mapped back onto 
Dunkl-Schr\"odinger form. While in general these issues must be solved on a case-by-case basis, they can be 
avoided entirely by restricting the mass function and the coordinate 
change as follows
\begin{eqnarray}
m(x) ~=~ p~x^q \qquad \qquad \qquad x(y) ~=~ \exp(y), \label{mxx}
\end{eqnarray}
where $p$ is an arbitrary constant, and $q$ is an even integer. Note that our coordinate change is invertible on the whole real axis. 
Before we continue, let us point out that the 
mass function in (\ref{mxx}) for $q \neq 0$ is either singular at the origin or vanishes there. While in most applications 
this is not a desired behavior, here we introduce the mass function in (\ref{mxx}) merely due to a 
specific mathematical property. More precisely, the effect of the settings (\ref{mxx}) on our equation (\ref{dsse1even}) 
is that the Dunkl parameters now appear as a constant term in the coefficient of the solution $\Phi$:
\begin{eqnarray}
\Phi''(y) \hspace{-.2cm} &+& \hspace{-.2cm}
\Bigg(-\frac{(q+1)^2}{4}+(q+1)~\delta~\nu-\nu^2+2~p~\exp\left[(q+2)~y \right] \left\{E- V_E\left[\exp(y)\right] \right\}
\Bigg)~\Phi(y) ~=~0. \nonumber \\ \label{sse0x}
\end{eqnarray}
We can rewrite this equation in the form
\begin{eqnarray}
\Phi''(y) +\left[\epsilon-U_E(y)\right]~\Phi(y) ~=~0, \label{sse1x}
\end{eqnarray}
introducing the abbreviations $\epsilon$ and $U_E$ that are given by
\begin{eqnarray}
\epsilon &=& (q+1)~\delta~\nu-\nu^2 \label{epsilon} \\[1ex]
U_E(y) &=& \frac{(q+1)^2}{4}-2~p~\exp\left[(q+2)~y \right] \left\{E- V_E\left[\exp(y)\right] \right\}. \label{ueve}
\end{eqnarray}
Therefore, we can now choose $\epsilon$ as the Darboux transformation parameter instead of the stationary energy. As a result, 
the transformed potential will not depend on the Dunkl parameters, and the transformed equation can be converted back to 
Dunkl-Schr\"odinger form in a straightforward manner. Application of the Darboux transformation (\ref{susy1}) and 
(\ref{susy2}) while replacing $U$ by $U_E$, we obtain the transformed solution and the associated potential as
\begin{eqnarray}
\hat{\Phi}(y) ~=~ \frac{W_{u_1,...,u_n,\Phi}(y)}{W_{u_1,...,u_n}(y)} \qquad \qquad \qquad 
\hat{U}_E(y) ~=~ U_E(y)- 2~\frac{d^2}{dy^2}~\log\left[
W_{u_1,...,u_n}(y)
\right], \nonumber
\end{eqnarray}
These functions enter in the transformed counterpart of (\ref{sse1x}) that reads
\begin{eqnarray}
\hat{\Phi}''(y) +\left[\epsilon-\hat{U}_E(y)\right]~\hat{\Phi}(y) ~=~0. \label{sse1xt}
\end{eqnarray}
Upon rewriting this equation in explicit form, we obtain the transformed partner of (\ref{sse0x}) as
\begin{eqnarray}
\hat{\Phi}''(y) \hspace{-.2cm} &+& \hspace{-.2cm}
\Bigg(-\frac{(q+1)^2}{4}+(q+1)~\delta~\nu-\nu^2+2~p~\exp\left[(q+2)~y \right] \left\{E- \hat{V}_E\left[\exp(y)\right] \right\}
\Bigg) \hat{\Phi}(y) ~=~0. \nonumber \\[1ex] \label{sse2xt}
\end{eqnarray}
Observe that the potential $\hat{V}_E$ can be obtained from the transformed version of (\ref{ueve}) that is given by
\begin{eqnarray}
\hat{U}_E(y) &=& \frac{(q+1)^2}{4}-2~p~\exp\left[(q+2)~y \right] \left\{E- \hat{V}_E\left[\exp(y)\right] \right\}. \nonumber
\end{eqnarray}
Solving with respect to $\hat{V}_E$ yields the result
\begin{eqnarray}
\hat{V}_E\left[\exp(y)\right] &=& E-\exp\left[-(q+2)~y \right] \left[\frac{(q+1)^2}{8~p}-\frac{\hat{U}_E(y)}{2~p}\right]. \label{vey}
\end{eqnarray}
Now that the Darboux transformation is complete, the remaining task consists in converting our equation (\ref{sse2xt}) back to 
Dunkl-Schr\"odinger form by reversing the point transformation (\ref{pct}). Recall that in the present case we made the 
settings (\ref{mxx}), such that the reversed point transformation reads
\begin{eqnarray}
\hat{\Psi}(x) &=& \sqrt{\frac{m(x)}{y'(x)}}~x^{-\nu}~\hat{\Phi}[y(x)] ~=~ \sqrt{p}~x^{\frac{1}{2}+\frac{q}{2}-\nu}~\hat{\Phi}\left[\log(x)\right], 
\label{rpct}
\end{eqnarray}
note that the function $y=y(x)=\log(x)$ stands for the inverse of $x=x(y)=\exp(y)$. Next, we rewrite the transformed potential 
(\ref{vey}) in terms of the variable $x$. This gives
\begin{eqnarray}
\hat{V}_E(x) &=& E- x^{-q-2} \left\{\frac{(q+1)^2}{8~p}-\frac{\hat{U}_E\left[\log(x)\right]}{2~p} \right\}. \label{vex}
\end{eqnarray}
Upon implementing the settings (\ref{rpct}) and (\ref{vex}) in our equation (\ref{sse2xt}), we obtain its explicit form as
\begin{eqnarray}
& & \hspace{-1cm} \frac{1}{2~m(x)}~\hat{\Psi}''(x)+
\Bigg[\hspace{-.1cm}
-\frac{m'(x)}{2~m(x)^2}+\frac{\nu}{m(x)~x}
\Bigg] \hat{\Psi}'(x)+\Bigg[-
\frac{\nu}{2~m(x)~x^2}-
\frac{\nu~m'(x)}{2~m(x)^2~x}+\frac{\nu~\delta}{2~m(x)~x^2}+ \nonumber \\[1ex]
& &\hspace{-1cm} +~\frac{\nu~\delta~m'(x)}{2~m(x)^2~x}+E-\hat{V}_E(x) \Bigg] \hat{\Psi}(x)~=~0, \nonumber
\end{eqnarray}
recall that the mass is restricted to the form shown in (\ref{mxx}). If the function $\hat{\Psi}$ has the correct 
parity as indicated by the value of $\delta$, then it is a solution of the transformed Dunkl-Schr\"odinger equation 
(\ref{sse0t}). In summary, we have established Darboux transformations between the Dunkl-Schr\"odinger partner 
equations (\ref{dsse0}) and (\ref{dsse0t}), provided the position-dependent mass function obeys the restriction (\ref{mxx}).

\subsection{Applications}
In order to illustrate the formalism developed in the previous paragraphs, we will now perform examples of second-order Darboux 
transformations with the goal of generating solvable Dunkl-Schr\"odinger equations with an energy-dependent potentials that 
admits solutions of bound state type. While in the first application the standard algorithm is used, the second application is devoted to the 
confluent algorithm. Our starting point for both applications is the Dunkl-Schr\"odinger equation (\ref{sse0}) that we will now equip 
with the following settings
\begin{eqnarray}
m(x) ~=~ \frac{1}{2} \qquad \qquad \qquad V_E(x) ~=~ \frac{x^2}{E}. \label{mv}
\end{eqnarray}
There are several motivations for choosing a constant mass function. Besides resulting in comparably simple calculations, 
our constant mass is a particular case of both (\ref{massgen}) and (\ref{mxx}) for the settings $p=1/2$, $q=0$. We will comment on 
different mass functions at the end of this paragraph. Now, upon substituting (\ref{mv}), the 
Dunkl-Schr\"odinger equation (\ref{sse0}) takes the specific form (\ref{mconst}). Furthermore, a constant mass implies 
even parity, such that we have $\mu=1$. 
In the next step we observe that the above choice for our mass function matches (\ref{mxx}) if we set $p=1/2$ and $q=0$. 
After incorporation of these settings into equation (\ref{mconst}), we obtain 
\begin{eqnarray}
\Psi''(x)+\frac{2~\nu}{x}
~\Psi'(x)+\Bigg(\frac{\delta~\nu-\nu}{x^2}+E-\frac{x^2}{E} \Bigg) \Psi(x)~=~0, \label{sseini}
\end{eqnarray}
which is a special case of (\ref{sse0x}). A particular solution to this equation can be given in the form
\begin{eqnarray}
\Psi(x) &=& \exp\left(\frac{x^2}{2 \sqrt{E}}\right) x^{\frac{1}{2}-\nu+\frac{1}{2}\sqrt{1-4\delta\nu+4\nu^2}}~
L_{-\frac{1}{2}+\frac{E^\frac{3}{2}}{4}-\frac{1}{4}\sqrt{1-4\delta\nu+4\nu^2}}^{\frac{1}{2}\sqrt{1-4\delta\nu+4\nu^2}}\left(\frac{x^2}{\sqrt{E}} 
\right), \label{solini}
\end{eqnarray}
note that $L$ stands for the associated Laguerre function \cite{abram}. Recall that the function (\ref{solini}) is a solution of the Dunkl-Schr\"odinger equation 
(\ref{sse0}) only if it has parity that matches the correct value of $\delta$. Let us now apply our point transformation (\ref{pct}) 
that takes (\ref{sseini}) into a form that is needed for applying our Darboux transformation. We choose
\begin{eqnarray}
\Phi(y) ~=~ \exp\left[\frac{1}{2}~y~(2~\nu-1)\right] \Psi\left[\exp(y)\right] \qquad \qquad \qquad x(y) ~=~ \exp(y). \label{pct1}
\end{eqnarray}
Observe that the left part is obtained from (\ref{pct}) by inserting the present settings (\ref{mv}) for $\mu=1$, $p=1/2$ and $q=0$. 
Upon inserting the explicit form (\ref{solini}), the function $\Phi$ is given by
\begin{eqnarray}
\Phi(y) &=& \exp\left[-\frac{1}{2~\sqrt{E}}~\exp(2~y)+\frac{y}{2}\sqrt{1-4~\delta~\nu+4~\nu^2}\right]
L_{-\frac{1}{2}+\frac{E^\frac{3}{2}}{4}-\frac{1}{4}\sqrt{1-4\delta\nu+4\nu^2}}^{\frac{1}{2}\sqrt{1-4\delta\nu+4\nu^2}}
\left[ \frac{\exp(2~y)}{\sqrt{E}} \right]. \nonumber \\[1ex] \label{phiini}
\end{eqnarray}
As such, it is a particular solution of the equation
\begin{eqnarray}
\Phi''(y)+\left[\delta~\nu-\nu^2-\frac{1}{4}+E~\exp(2~y)-\frac{\exp(4~y)}{E} \right] \Phi(y) &=& 0, \label{sse1}
\end{eqnarray}
which is a special case of (\ref{sse1x}). As expected, the Dunkl parameters $\delta$ and $\nu$ appear as a constant term 
that is independent of the variable $y$. Let us follow the definitions from (\ref{epsilon}) and (\ref{ueve}) by setting
\begin{eqnarray}
\epsilon ~=~ \delta~\nu-\nu^2 \qquad \qquad \qquad U_E(y) ~=~ \frac{1}{4}-E~\exp(2~y)+\exp(4~y). \label{epsw}
\end{eqnarray}
These definitions can be implemented in equation (\ref{sse1}) that takes the Schr\"odinger-like form (\ref{sse1x}). Before we 
proceed with the Darboux transformation, let us recall a remark we made in section 4 about conditions (\ref{con}) and (\ref{con0}). 
According to the remark, the same function $U_E$ can be constructed from various choices of the parameters $m$, $V_E$ and 
the coordinate change $x$. Let us now demonstrate this using the $U_E$ given in (\ref{epsw}), recall that we obtained the latter 
function through the settings (\ref{mv}). If we replace these settings by
\begin{eqnarray}
m(x) ~=~ \frac{1}{2}~x^2 \qquad \qquad \qquad V_E(x) ~=~ E+\frac{1}{E}-\frac{E}{x^2}-\frac{2}{x^4}, \label{mvx}
\end{eqnarray}
and perform our point transformation (\ref{pct1}), we obtain the equation
\begin{eqnarray}
\Phi''(y)+\left[3~\delta~\nu-\nu^2-\frac{1}{4}+E~\exp(2~y)-\frac{\exp(4~y)}{E} \right] \Phi(y) &=& 0. \label{sse10}
\end{eqnarray}
We observe that up to a numerical factor, this equation has the same form as its counterpart (\ref{sse1}). We can match both 
forms by formally redefining the parameter $\nu$ in (\ref{sse10}) by means of new parameters $\bar{\delta}$ and $\bar{\nu}$ as
\begin{eqnarray}
\nu &=& \frac{3~\delta}{2}+\frac{1}{2}~\sqrt{9-4~\bar{\delta}~\bar{\nu}+4~\bar{\nu}^2}. \nonumber
\end{eqnarray}
Substitution renders our equation (\ref{sse10}) in the form
\begin{eqnarray}
\Phi''(y)+\left[\bar{\delta}~\bar{\nu}-\bar{\nu}^2-\frac{1}{4}+E~\exp(2~y)-\frac{\exp(4~y)}{E} \right] \Phi(y) &=& 0, \nonumber
\end{eqnarray}
which coincides with (\ref{sse1}) up to parameter naming. Hence, the settings (\ref{mv}) for a constant mass system and 
(\ref{mvx}) for a position-dependent mass system yield the same equation. We can therefore consider the results obtained by 
application of the Darboux transformation in the subsequent paragraphs as valid for both the constant mass case as well as 
certain position-dependent mass scenarios. We are now 
ready to perform our Darboux transformation, such that we must distinguish between the standard and the confluent 
algorithm.

\subsubsection{Standard Darboux algorithm}
\begin{figure}[H]
\begin{center}
\epsfig{file=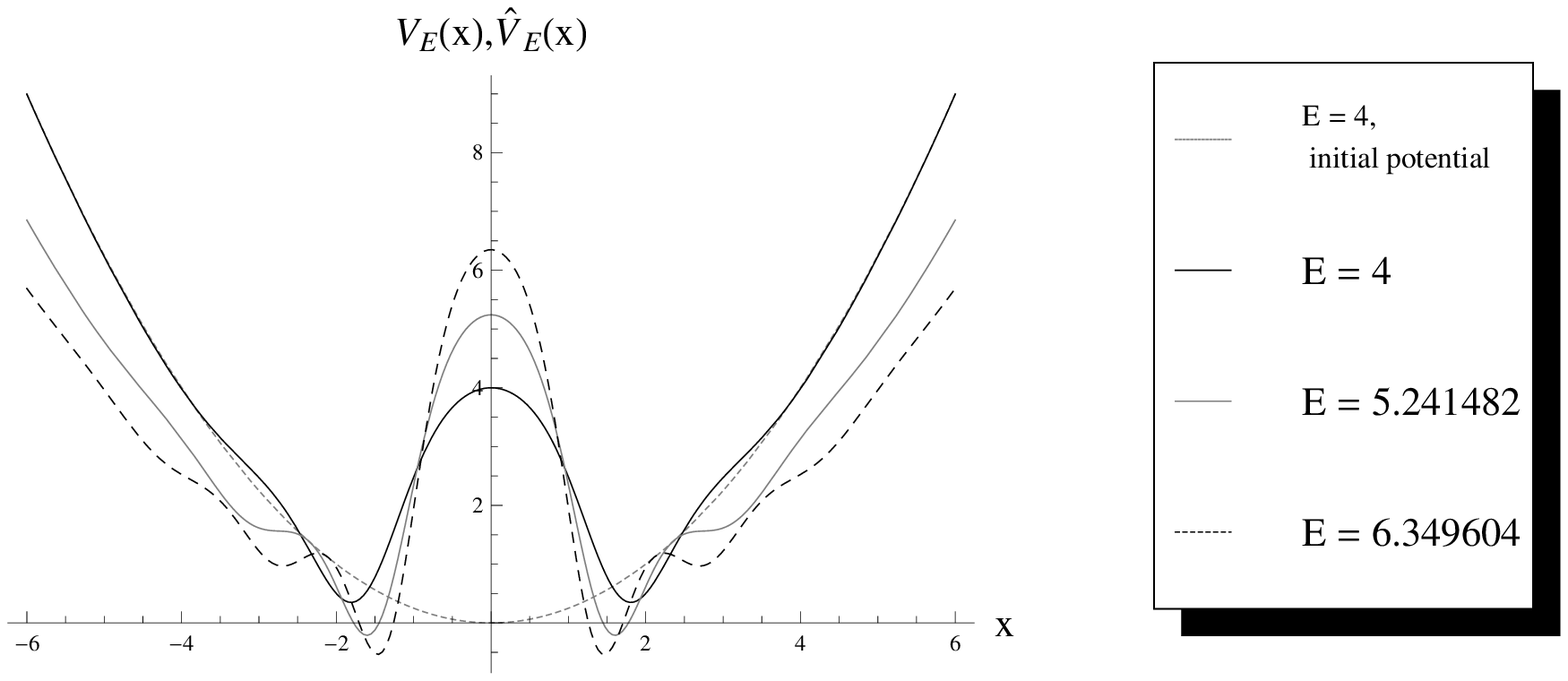,width=16cm}
\caption{Graphs of the initial potential from (\ref{mv}) and transformed counterparts (\ref{hatv}) for the settings (\ref{mv}) and the 
transformation functions (\ref{u10}) and (\ref{u20}).} 
\label{fig3}
\end{center}
\end{figure}

\begin{figure}[H]
\begin{center}
\epsfig{file=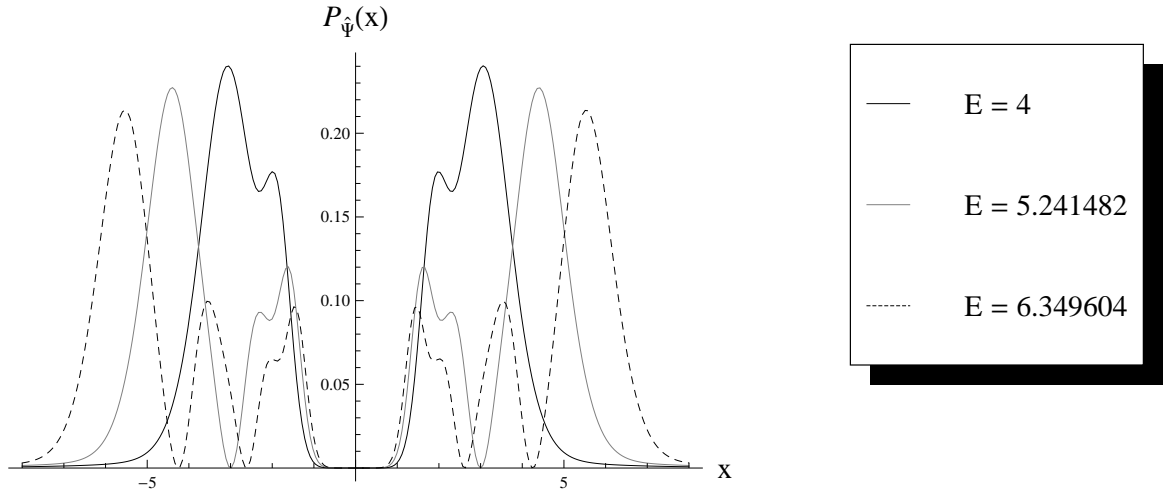,width=16cm}
\caption{Graphs of the normalized transformed probability density for the parameter settings $\delta=-1$ and $\nu=5/2$.} 
\label{fig4}
\end{center}
\end{figure}

\begin{figure}[H]
\begin{center}
\epsfig{file=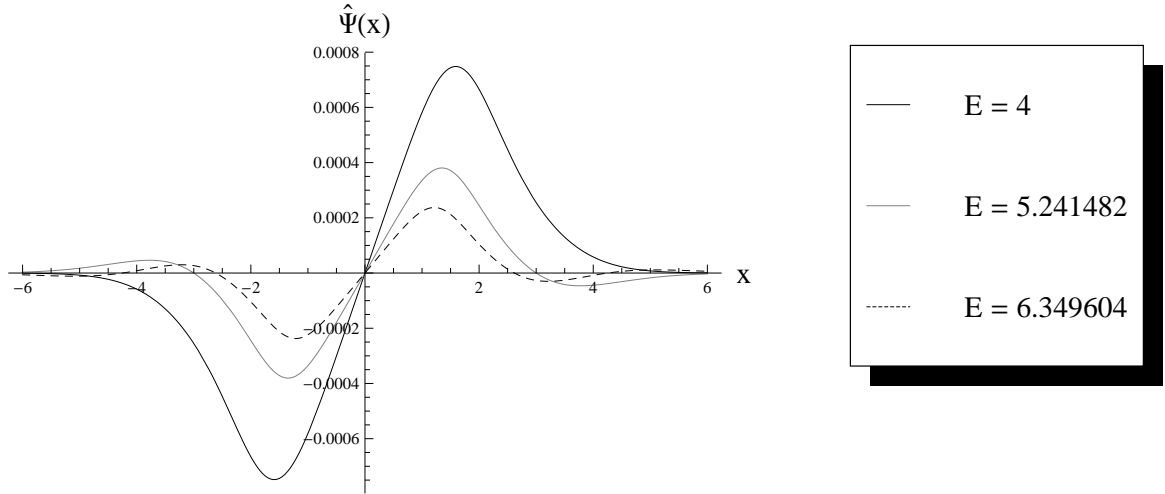,width=16cm}
\caption{Graphs of the normalized transformed probability density for the parameter settings $\delta=-1$ and $\nu=5/2$.} 
\label{fig5}
\end{center}
\end{figure}

\begin{figure}[H]
\begin{center}
\epsfig{file=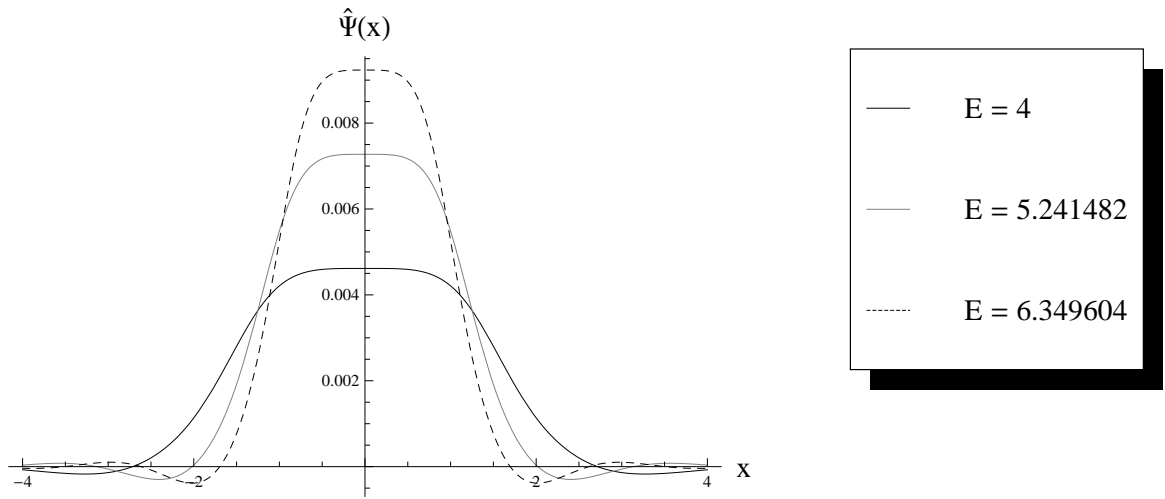,width=16cm}
\caption{Graphs of the normalized transformed probability density for the parameter settings $\delta=1$ and $\nu=7/2$.} 
\label{fig6}
\end{center}
\end{figure}

This algorithm requires that we provide two transformation functions $u_1,u_2$ that solve (\ref{sse1x}) for the present settings. 
We choose these functions as special cases of our soution (\ref{phiini}) as
\begin{eqnarray}
u_1(y) ~=~ \Phi(y)_{\epsilon=\epsilon_1} \qquad \qquad \qquad u_2(y) ~=~ \Phi(y)_{\epsilon=\epsilon_2}. \label{u12}
\end{eqnarray}
In order to keep our calculations transparent, we want these functions to take a relatively simple form. This can be achieved by selecting the parameters 
$\epsilon_1$ and $\epsilon_2$ as
\begin{eqnarray}
\epsilon_1 ~=~ \frac{1}{4} \qquad \qquad \qquad \epsilon_2~=~-\frac{3}{4}. \nonumber
\end{eqnarray}
Substitution of these values into our transformation functions (\ref{u12}) gives after simplifying and collecting terms
\begin{eqnarray}
u_1(y) &=& \exp\left[-\frac{1}{2~\sqrt{E}}~\exp(2~y)\right] L_{\frac{E^\frac{3}{2}}{4}-\frac{1}{2}}\left[\frac{\exp(2~y)}{\sqrt{E}}\right] \label{u10} \\[1ex]
u_2(y) &=& \exp\left[y-\frac{1}{2~\sqrt{E}}~\exp(2~y)\right] L_{\frac{E^\frac{3}{2}}{4}-1}^{1}\left[\frac{\exp(2~y)}{\sqrt{E}}\right]. \label{u20}
\end{eqnarray}
The transformed solution $\hat{\Phi}$ is obtained from the rule (\ref{susy1}) for the case $n=2$, that is, we have
\begin{eqnarray}
\hat{\Phi}(y) &=& \frac{W_{u_1,u_2,\Phi}(y)}{W_{u_1,u_2}(y)}. \label{hatphi}
\end{eqnarray}
Furthermore, the transformed potential $\hat{U}_E$ reads
\begin{eqnarray}
\hat{U}_E(y) &=& U_E(y)-2~\frac{d^2}{dx^2} \log\left[W_{u_1,u_2}(y)\right], \label{hatw}
\end{eqnarray}
recall that $U_E$ is given in (\ref{epsw}). We omit to state the explicit form of the functions (\ref{hatphi}) and (\ref{hatw}) due to their 
length, but we will give particular cases below after having reverted the point transformation. The solution (\ref{hatphi}) and 
the potential (\ref{hatw}) enter in equation (\ref{sse1xt}), which in the next step we map back to Dunkl-Schr\"odinger form 
by the inverted version of (\ref{pct1}). It is given by
\begin{eqnarray}
\hat{\Psi}(x) ~=~x^{\frac{1}{2}-\nu}~\hat{\Phi}\left[\log(x)\right] \qquad \qquad \qquad y(x) ~=~ \log(x). \label{hatpsi}
\end{eqnarray}
As mentioned above, the explicit form of this function is very large, such that we restrict ourselves to stating a special case here for 
the settings $|\nu=5/2$, $\delta=1$, and $E=6$. Substitution yields
\begin{eqnarray}
\hat{\Psi}(x)_{|\nu=5/2,\delta=-1,E=4} ~=~ \frac{\exp\left(-\frac{x^2}{2} \right) I_0\left(\frac{x^2}{2}\right)}
{(24-6~x^2+x^4)~I_0\left(\frac{x^2}{2}\right)-x^2~(x^2-4)~I_1\left(\frac{x^2}{2}\right)}, \label{solx}
\end{eqnarray}
where $I$ stands for the modified Bessel function of the first kind \cite{abram}. This solution represents a bound state for our 
transformed system, as we will demonstrate below. Next, we compute the transformed potential $\hat{V}_E$ from 
(\ref{vex}) by inserting the present settings $p=1/2$ and $q=0$, along with (\ref{hatw}). We obtain
\begin{eqnarray}
\hat{V}_E(x) &=& E-\frac{1}{4~x^2}+\frac{1}{x^2}~\hat{U}_E\left[\log(x)\right]. \label{hatv}
\end{eqnarray}
Note that $\hat{V}_E$ depends on the energy $E$. As in case of the transformed solution, we will give a particular 
potential (\ref{hatv}) for $E=4$. Evaluation yields
\begin{eqnarray}
\hat{V}_4(x) &=&\Bigg\{4~\Bigg[(24-6~x^2+x^4)~I_0\left(\frac{x^2}{2}\right)-x^2~(x^2-4)~I_1\left(\frac{x^2}{2}\right) \Bigg]^2
\Bigg\}^{-1} \times \nonumber \\[1ex]
&\times&
\Bigg[
(9216-7488~x^2+1920~x^4-180~x^6+4~x^8+x^{10})~I_0\left(\frac{x^2}{2}\right)^2- \nonumber \\[1ex]
&-& 2~x^2~(x-2)~(x+2)~(x^2+16)~(24-6~x^2+x^4)~
I_0\left(\frac{x^2}{2}\right)I_1\left(\frac{x^2}{2}\right)+ \nonumber \\[1ex]
&+& x^2~(x^2-4)^2~(72+16~x^2+x^4)~I_1\left(\frac{x^2}{2}\right)^2\Bigg]. \nonumber
\end{eqnarray}
Figure \ref{fig3} shows a graph of this function, along with the initial potential from (\ref{mv}) and other transformed special 
cases for specific choices of the stationary energy.
 \noindent
The solution (\ref{hatpsi}) and potential (\ref{hatv}) enter in the transformed Dunkl-Schr\"odinger equation that is the partner 
to (\ref{sseini}). It reads
\begin{eqnarray}
\hat{\Psi}''(x)+\frac{2~\nu}{x}
~\hat{\Psi}'(x)+\Bigg[\frac{\delta~\nu-\nu}{x^2}+E-\hat{V}_E(x)\Bigg] \hat{\Psi}(x)~=~0. \label{ssett}
\end{eqnarray}
Let us now recall that the function (\ref{hatpsi}) is an admissible solution for this equation only if its parity matches the 
correct value of $\delta$. Inspection of the explicit form of (\ref{hatpsi}) reveals that parity is determined by a single monomial 
term. We have 
\begin{eqnarray}
\hat{\Psi}(x) &\stackrel{x<<1}{=}& x^{\frac{1}{2}-\nu+\frac{1}{2} \sqrt{1-4\delta\nu+4\nu^2}}, \label{expo}
\end{eqnarray}
where irrelevant multiplicative constants were discarded. The exponent in (\ref{expo}) must be an odd integer if $\delta=-1$, and it 
must be an even integer for $\delta=1$. Let us evaluate the exponent for the first case. We find
\begin{eqnarray}
\frac{1}{2}-\nu+\frac{1}{2} ~\sqrt{1+4~\nu+4~\nu^2} &=& \frac{1}{2}-\nu+\frac{1}{2}~|1+2~\nu| ~=~ 1, \nonumber 
\end{eqnarray}
note that in the last step we used the restriction $\nu>-1/2$. In the second case $\delta=1$, the exponent (\ref{expo}) takes the form
\begin{eqnarray}
\frac{1}{2}-\nu+\frac{1}{2}~ \sqrt{1-4~\nu+4~\nu^2} &=& \frac{1}{2}-\nu+\frac{1}{2}~|-1+2~\nu| \nonumber \\[1ex]
&=& \left\{
\begin{array}{lll}
1-2~\nu & \mbox{if} & \nu < \frac{1}{2} \\[1ex]
0 & \mbox{if} & \nu \geq \frac{1}{2}
\end{array}
\right\}. \nonumber
\end{eqnarray}
Since $\nu>-1/2$, the exponent cannot attain any even integer value for $\nu<1/2$. Thus, we must have $\nu \geq 1/2$, in which 
case the exponent vanishes. Consequently, the correct parity of our transformed solution (\ref{hatpsi}) is established if we 
choose $\nu \geq 1/2$. In the next step we will establish normalizability of our solutions (\ref{hatpsi}) for certain values of the 
energy. To this end, we recall that the modified norm of a solution to the Dunkl-Schr\"odinger equation (\ref{sse0}) 
is given by (\ref{norm}). In order to be meaningful, the integral this modified norm must exist, and the integrand must be 
nonnegative. Existence of the integral can be achieved by choosing the stationary energy such that the first argument of the 
Laguerre function in (\ref{solini}) becomes a nonnegative integer. This condition reads
\begin{eqnarray}
-\frac{1}{2}+\frac{E^\frac{3}{2}}{4}-\frac{1}{4}\sqrt{1-4~\delta~\nu+4~\nu^2} &=& n, \nonumber
\end{eqnarray}
where $n$ is a nonnegative integer. Solving with respect to $E$ gives
\begin{eqnarray}
E &=& \left(4~n+2+\sqrt{1-4~\delta~\nu+4~\nu^2}\right)^\frac{2}{3}. \label{ene1}
\end{eqnarray}
As a consequence of this choice for the stationary energy, the Laguerre function will degenerate to a 
polynomial, which renders (\ref{solini}) bounded on the whole real line. This property persists under our Darboux transformation, 
such that the transformed solution (\ref{hatpsi}) is also bounded on the whole real line. This extends to the whole integrand in 
(\ref{norm}), such that the integral exists. In addition we have
\begin{eqnarray}
1-\frac{\partial V_E(x)}{\partial E} &\geq& 0. \nonumber
\end{eqnarray}
We do not verify these conditions explicitly here due to the length of the involved expressions. Instead we show graphs of 
normalized probability densities associated with our transformed solution (\ref{hatpsi}) for different parameter values in 
figure \ref{fig4}. 
 \noindent
Recall that the probability density for a solution $\Psi$ is given by (\ref{prob}). Inspection of the figure indicates that the 
probability densities are nonnegative functions. Note that the energy values chosen in the figures \ref{fig3} and \ref{fig4} 
are taken from (\ref{ene1}) for $n=0,1,2$. Since we know now that our transformed solution (\ref{hatpsi}) represents bound states 
if the stationary energy is chosen from (\ref{ene1}), let us present graphs of particular cases. Figure \ref{fig5} and figure \ref{fig6} 
show graphs of the solutions (\ref{hatpsi}) for the odd-parity case and the even-parity case, respectively.
 \noindent
 \noindent
Observe that one of the graphs shown in figure \ref{fig5} pertains to the function (\ref{solx}).

\subsubsection{Confluent Darboux algorithm}
\begin{figure}[H]
\begin{center}
\epsfig{file=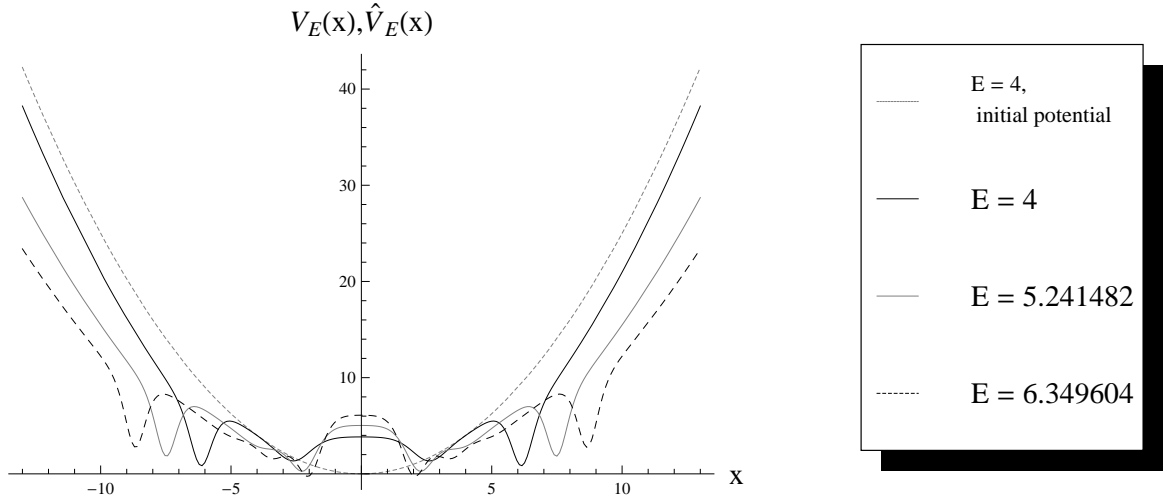,width=16cm}
\caption{Graphs of the initial potential from (\ref{mv}) and transformed counterparts (\ref{hatv}) for the settings (\ref{mv}) and the 
transformation functions (\ref{u1c}) and (\ref{u2c}).} 
\label{fig7}
\end{center}
\end{figure}

We go back to our initial equation (\ref{sse1}) that we consider to be in the form (\ref{sse1x}) due to our settings 
(\ref{epsw}). In contrast to its standard counterpart, the confluent algorithm of our Darboux transformations requires 
transformation functions that solve the system (\ref{eq1}), (\ref{eqn}). In the present case of a second-order transformation 
the latter system reads
\begin{eqnarray}
u_1''(y)+\left[\epsilon_1-U_E(y) \right] u_1(y) &=& 0 \label{eq1x} \\[1ex]
u_2''(y)+\left[\epsilon_1-U_E(y)\right]~u_2(y) &=&- u_{1}(y), \label{eq2x}
\end{eqnarray}
recall that $U_E$ is given in (\ref{epsw}). In order to solve the system (\ref{eq1x}), (\ref{eq2x}), we observe that the first equation 
is formally the same as (\ref{sse1}), such that a solution $u_1$ can be taken from (\ref{phiini}). After inserting the definition of $\epsilon$ 
in (\ref{epsw}) and replacing $\epsilon$ by $\epsilon_1$, we find
\begin{eqnarray}
u_1(y) &=& \exp\left[-\frac{1}{2~\sqrt{E}}~\exp(2~y)+\frac{y}{2}\sqrt{1-4~\epsilon_1}\right]
L_{-\frac{1}{2}+\frac{E^\frac{3}{2}}{4}-\frac{1}{4}\sqrt{1-4 \epsilon_1}}^{\frac{1}{2}\sqrt{1-4 \epsilon_1}}
\left[ \frac{\exp(2~y)}{\sqrt{E}} \right]. \label{u1c0}
\end{eqnarray}
In the present application we choose the parameter $\epsilon_1$ as
\begin{eqnarray}
\epsilon_1 &=& -2. \nonumber
\end{eqnarray}
Upon substitution into (\ref{u1c0}), we obtain the explicit form
\begin{eqnarray}
u_1(y) &=& \exp\left[-\frac{1}{2~\sqrt{E}}~\exp(2~y)+\frac{3~y}{2}\right]
L_{\frac{E^\frac{3}{2}}{4}-\frac{5}{4}}^{\frac{3}{2}}
\left[ \frac{\exp(2~y)}{\sqrt{E}} \right]. \label{u1c}
\end{eqnarray}
In the next step we need to find the remaining transformation function $u_2$ by solving equation (\ref{eq2x}), which can be done by 
applying integral or differential formulas \cite{xbatconfluent}. According to these formulas, the simplest way of constructing a 
particular solution to (\ref{eq2x}) is given by 
\begin{eqnarray}
u_2(y) &=& \frac{\partial \Phi(y)}{\partial \epsilon}_{\mid \epsilon = \epsilon_1}, \label{u2c}
\end{eqnarray}
where as before $\Phi$ was taken from (\ref{phiini}). We omit to state the explicit form of the resulting transformation function 
$u_2$ after application of the derivative. From this point on the confluent algorithm of the Darboux transformation proceeds in the 
same way as the standard case. Substitution of the functions (\ref{u1c}), (\ref{u1c}), (\ref{phiini}) into (\ref{hatphi}) and (\ref{hatw}) 
yields the transformed solution and potential, respectively. After reversing our point transformation, we obtain a solution 
(\ref{hatpsi}) and an associated potential (\ref{hatv}) for equation (\ref{ssett}). If the latter solution has the correct parity as indicated 
by the parameter $\delta$, then it solves the transformed Dunkl-Schr\"odinger equation (\ref{sse0t}) for the present settings in 
(\ref{mv}). Due to the parametric derivatives, the transformed solution and potential do not allow for simplification, such that their 
explicit forms are long and involved. As a consequence, it becomes difficult to analyze the probability density associated with the 
solutions in explicit form. For this reason, we restrict ourselves here to show graphs of the transformed potential only, see 
figure \ref{fig7}.
 \noindent

\section{Concluding remarks}
When applying solution-generating methods like point or Darboux transformations, the presence of an energy-dependent 
potential imposes an additional constraint on the transformed system, in particular when looking for bound states. This constraint 
arises from the norm that involves the parametric derivative of the initial system's potential. In general, normalizability is not 
preserved by either transformation. However, if we assume that the coordinate change $x=x(y)$ does not depend on the energy, then 
we can derive the following relation from (\ref{con}):
\begin{eqnarray}
1-\frac{\partial U_E(y)}{\partial E} &=& 2~m[x(y)]~[x'(y)]^2 \left[1-\frac{\partial V_E(y)}{\partial E}\right]. \nonumber
\end{eqnarray}
We observe that if the mass is nonnegative and the coordinate change is real-valued, then normalizability is preserved by 
the point transformation (\ref{pct}). For the Darboux transformation an analogous statement cannot be obtained in such a simple 
manner because the transformed potential depends on the transformation functions that can exhibit a complicated dependence 
on the energy. A further comment on the Darboux transformations and the restriction (\ref{mxx}) is in order. This restriction on 
the mass can be avoided by assigning the parity parameter $\delta$ a fixed value of $-1$ or $1$. Since in this case the 
transformed potential does not depend on this parameter anymore, the restriction (\ref{mxx}) is not necessary. On the other hand, the 
Darboux-transformed equation must then be cast in a form that allows mapping it back to Dunkl-Schr\"odinger form, and parity of the 
transformed solutions must be established.

\paragraph{Data availability statement.} The data that supports the findings of this study are available within the article.

\end{sloppypar}

\end{document}